\documentstyle [12pt] {article}
\title{
    New Projects of Crystal Extraction at IHEP 70 GeV Accelerator
       }
\author{
A.G.Afonin, \underbar{V.M.Biryukov},
V.N.Chepegin, Y.A.Chesnokov,\\ V.I.Kotov,
V.I.Terekhov, E.F.Troyanov, Yu.S. Fedotov\\
{\em Institute for High Energy Physics, Protvino, Russia}\\
Yu.M.Ivanov,
{\em Nuclear Physics Institute, St.Petersburg, Russia}\\
W.Scandale,
{\em CERN, Geneva};
M.B.H.Breese,
{\em University of Surrey, UK}
}
\date{Talk given at PAC 1999 (New York)}
\begin{document}
\maketitle

\begin{abstract}
Using channeling in a 5-mm crystal with bending angle
of 0.65 mrad, a record high efficiency, over 60\%,
of particle extraction from
accelerator was achieved.
The extracted beam intensity was up to 5.2$\times$10$^{11}$ protons per
spill of $\sim$0.5 s duration.
Also, the first proof-of-principle experiment on 'crystal collimation'
was performed where crystal - serving as a scraper -
has reduced the radiation level
in the accelerator by a factor of two.
The measurements agree with Monte Carlo predictions.
\end{abstract}

\section{Introduction}

Crystal can channel a charged particle
if it comes within so-called critical angle $\theta_c$,
about $\pm$5 $\mu$rad/$\sqrt{p(TeV)}$ in silicon.
This restricts  crystal efficiency in divergent beams.
However, it's been argued theoretically\cite{theory}
that a breakthrough in crystal efficiency can be due to
multiple character of particle encounters with a crystal
installed in a circulating beam,
where particle may scatter in inefficient encounters
and have new chances on later turns.
The importance of multi-pass mechanism of extraction
was first shown in CERN SPS experiment\cite{cern}.
To benefit from the "multi-pass" channeling,
the crystal must be short enough to reduce beam losses
in multiple encounters with it.

The Protvino experiment on slow extraction of 70 GeV protons
from IHEP accelerator employs, as a primary element of the extraction
system, a very short bent silicon crystal.
To gain extraction efficiency from
an increased number of proton encounters with the crystal,
it is necessary to use crystals as short as possible.

Bending a short crystal to be installed in the accelerator
vacuum chamber is not easy.
The first crystal Si(111) was performed as a short plate of a big height,
0.5$\times$40$\times$7 mm$^3$ (thickness, height, and length along the beam
direction, respectively).
It was bent transversally with a metal holder which had a hole of 20 mm size
for beam passage, and gave the channeled protons a deflection of 1.7 mrad.
Despite an angular distortion (a "twist") in that design,
encouraging results on beam extraction were obtained in our first run
in December 1997, Figure 1.
The peak extraction efficiency reached about 20\% and
the extracted beam intensity was up to 1.9$\times$10$^{11}$ \cite{iheprep}.
Here and later on in the paper, the extraction efficiency is defined
as the ratio of the extracted beam intensity as measured in the external
beamline to all the beam loss in the accelerator.

To further increase the extraction efficiency,
further crystals (without twist)
were made from a monolithic Si piece in a shape of
"O" at the Petersburg Nuclear Physics Institute,
as described in Ref. \cite{plb}.
The crystals Si(110) used in our recent runs had
the length along the beam direction of only 5 mm.
The bent part of the crystal was just 3 mm long,
and the straight ends were 1 mm each.

Such a crystal, with bending angle of 1.5 mrad, was successfully
tested in March 1998 and has shown extraction efficiencies over 40\%
\cite{plb}. In the mean time we have changed the crystal location
in order to use another septum magnet (with partition thickness
of 2.5 mm instead of 8 mm as in the old scheme)
where a smaller bending angle is required from a crystal.
This change was also motivated by the intention to test
even shorter crystals (two of them, 2.5 and 3.0 mm long,
are already prepared for the tests).
The crystal used in this location was new, but of the same design
and dimensions as earlier described\cite{plb}.
The bending angle used in our recent run was 0.65 mrad.
Below we consider the results of the studies with this crystal.

\section{Experiment}

The general schematics of beam extraction by a crystal is shown
in Ref.\cite{plb}.
As the small angles of deflection are insufficient for a direct extraction
of the beam from the accelerator,
a crystal served as a primary element in
the existing scheme of slow extraction.
Crystal was placed in straight section 106 of the accelerator
upstream of a septum-magnet of slow-extraction system.
The accuracy of the crystal horizontal and angular
translations was 0.1 mm and 13.5 $\mu$rad, respectively.
The horizontal emittance of the circulating proton beam
was about 2$\pi$ mm$\times$mrad, and
the beam divergence at the crystal location was 0.6 mrad.
A local distortion of the orbit by means of bump windings
in magnets moved the beam slowly toward the crystal.
To obtain a uniform rate of the beam at crystal, a monitor
for close loop operation
based on a photomultiplier with scintillator was used
to automatically adjust the orbit distortion.
We used also function generator to control current in bump windings.

The beam deflection to the septum and its transmission
through the beam line of extraction were supervised with a complex system of
beam diagnostics, including TV system, loss monitors, profilometers,
intensity monitors\cite{plb}.
All the diagnostics devices were firstly tested in
fast-extraction mode and calibrated with beam transformers.
The background conditions were periodically measured
with and without crystal.
According to the measurements, the fraction of background particles
(e.g. elastically scattered protons)
together with the apparatus noise did not exceed 4\% of the
useful signal level. This background was subtracted from
the efficiency figures shown in the paper.
The fraction of the beam directed to the crystal
was defined as the difference
between the measurements of the circulating beam intensity
done with beam transformers
before and after the beam extraction,
with the systematic error of 1\%.
The extraction efficiency
was evaluated in every cycle of acceleration.

\section{Results}

The accelerator beam intensity during the experiment
was about 1.3$\times$10$^{12}$ protons per cycle.
The fraction of the circulating beam incident on the crystal
$\Delta$I was varied from 20 to 90\%.
The spill duration of the channeled beam in the feedback regime was
on the order of 0.5 s.
The plateau of the IHEP U-70 accelerator magnet cycle is 2 s long
while the overall cycle of the machine is 9.6 s.
Figure 1 shows the efficiency of extraction averaged over the spill,
as measured in our three experiments of 1997-98.
In the last one, the efficiency was about 50\% even when all the
accelerator beam was directed onto the crystal.
The spill-averaged efficiency figures were reproducible
with 1\% accuracy from run to run.
The dependence of the extracted beam intensity
on orientation of the crystal was about the same as in Ref.\cite{plb}
and not shown here.
The highest intensity of the extracted beam, for 1.15$\times$10$^{12}$
protons incident at the crystal in a cycle, was equal to
5.2$\times$10$^{11}$.

\begin{figure}
	\begin{center}
\setlength{\unitlength}{.65mm}
\begin{picture}(110,120)(0,-3)
\thicklines
\linethickness{.1mm}
\multiput(0,20)(2,0){50}{\line(1,0){1}}
\multiput(0,40)(2,0){50}{\line(1,0){1}}
\multiput(0,60)(2,0){50}{\line(1,0){1}}
\multiput(0,80)(2,0){50}{\line(1,0){1}}
\multiput(20,0)(0,2){50}{\line(0,1){1}}
\multiput(40,0)(0,2){50}{\line(0,1){1}}
\multiput(60,0)(0,2){50}{\line(0,1){1}}
\multiput(80,0)(0,2){50}{\line(0,1){1}}
\linethickness{.4mm}

\put(     90. ,25.5)  {\line(1,0){4}}
\put(     92. ,23.5)  {\line(0,1){4}}
\put(     59.5 ,31.7)  {\line(1,0){4}}
\put(     61.5 ,29.7)  {\line(0,1){4}}
\put(     39. ,34.)  {\line(1,0){4}}
\put(     41. ,32.)  {\line(0,1){4}}
\put(     21. ,41.6)  {\line(1,0){4}}
\put(     23. ,39.6)  {\line(0,1){4}}
\put(     14. ,38.)  {\line(1,0){4}}
\put(     16. ,36.)  {\line(0,1){4}}


\put(     89. ,45.)  {\circle*{3}}
\put(     62. ,50.)  {\circle*{3}}
\put(     41  ,48)  {\circle*{3}}
\put(     21  ,48)  {\circle*{3}}

\put(     98. ,13.)  {\circle {3}}
\put(     70. ,14)  {\circle {3}}
\put(     40. ,12.5)  {\circle {3}}
\put(     20. ,10)  {\circle {3}}



\linethickness{.2mm}
\put(0,0) {\line(1,0){100}}
\put(0,0) {\line(0,1){100}}
\put(0,100) {\line(1,0){100}}
\put(100,0){\line(0,1){100}}
\multiput(25,0)(25,0){3}{\line(0,1){2.5}}
\multiput(5,0)(5,0){20}{\line(0,1){1.4}}
\multiput(0,20)(0,20){4}{\line(1,0){2.5}}
\multiput(0,4)(0,4){25}{\line(1,0){1.4}}
\multiput(100,20)(0,20){4}{\line(-1,0){2.5}}
\multiput(100,4)(0,4){25}{\line(-1,0){1.4}}
\put(24,3){\makebox(1,1)[b]{25}}
\put(49,3){\makebox(1,1)[b]{50}}
\put(74,3){\makebox(1,1)[b]{75}}
\put(-9,20){\makebox(1,.5)[l]{20}}
\put(-9,40){\makebox(1,.5)[l]{40}}
\put(-9,60){\makebox(1,.5)[l]{60}}
\put(-9,80){\makebox(1,.5)[l]{80}}
\put(-9,100){\makebox(1,.5)[l]{100}}

\put(3,105){\large F(\%)}
\put(80,-10){\large $\Delta$I(\%)}

\end{picture}
	\end{center}
	\caption{
	Spill-averaged efficiency of extraction as measured with
	 5-mm crystal 0.65 mrad bent ($\bullet$), December 1998;
	 5-mm crystal 1.5 mrad bent ($+$), March 1998;
	 7-mm twisted crystal 1.7 mrad bent (o), December 1997;
	plotted against the beam fraction taken from the accelerator.
}
	\end{figure}
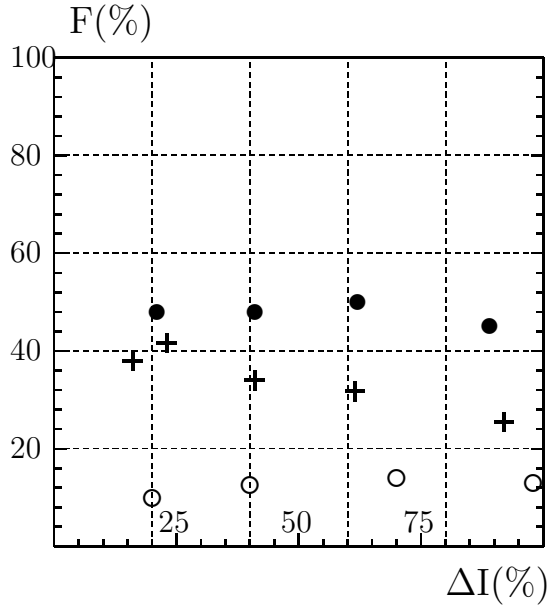

As the beam moves radially toward the crystal,
the proton incidence angle drifts at the crystal.
For this reason the extraction efficiency varies in time during the spill
(Fig. 2), especially for a large beam fraction used.
The peak extraction efficiency in a spill was always greater
than 60\%.
Figure 2 shows also a minor 'satellite peak',
present in some cases and unexplained yet.
The absolute extraction efficiency as obtained in
our Monte Carlo simulations agree with the measurements
to accuracy of about 5\% for spill-averaged figures.
The time dependence calculations (Figure 2)
are more sensitive to the details of beam direction methods.

\begin{figure}
	\begin{center}
\setlength{\unitlength}{.7mm}
\begin{picture}(110,120)(-5,-5)
\thicklines
\linethickness{.2mm}
\put(     6. ,17.1)  {\circle{2.5}}
\put(     9. ,22.5)  {\circle{2.5}}
\put(     12.9 ,28.9)  {\circle{2.5}}
\put(     19. ,36.9)  {\circle{2.5}}
\put(     27. ,46.7)  {\circle{2.5}}
\put(     38, 58.)  {\circle{2.5}}
\put(     44.5, 57)  {\circle{2.5}}
\put(     51,  46.6)  {\circle{2.5}}
\put(     59, 41.3)  {\circle{2.5}}
\put(     67,  39.8)  {\circle{2.5}}
\put(     76, 36.3)  {\circle{2.5}}
\put(     85,  31.7)  {\circle{2.5}}

\put(     5. ,28.)  {\circle*{2.5}}
\put(     6. ,35.)  {\circle*{2.5}}
\put(     7. ,41.)  {\circle*{2.5}}
\put(     8. ,48.)  {\circle*{2.5}}
\put(     12. ,53.)  {\circle*{2.5}}
\put(     16. ,59.)  {\circle*{2.5}}
\put(     20. ,64.)  {\circle*{2.5}}
\put(     26, 70.)  {\circle*{2.5}}
\put(     35, 73.)  {\circle*{2.5}}
\put(     44, 70.)  {\circle*{2.5}}
\put(     50, 64)  {\circle*{2.5}}
\put(     55, 60)  {\circle*{2.5}}
\put(     60, 55)  {\circle*{2.5}}
\put(     65, 51)  {\circle*{2.5}}
\put(     65.75, 54.5)  {\circle*{2.5}}
\put(     66.5, 58)  {\circle*{2.5}}
\put(     67, 55)  {\circle*{2.5}}
\put(     67.5, 51)  {\circle*{2.5}}
\put(     68, 48)  {\circle*{2.5}}
\put(     71, 42.5)  {\circle*{2.5}}
\put(     74, 37)  {\circle*{2.5}}
\put(     77, 31.5)  {\circle*{2.5}}
\put(     80, 26)  {\circle*{2.5}}
\put(     85, 21)  {\circle*{2.5}}
\put(     90, 16)  {\circle*{2.5}}

\linethickness{.2mm}
\put(0,0) {\line(1,0){100}}
\put(0,0) {\line(0,1){100}}
\put(0,100) {\line(1,0){100}}
\put(100,0){\line(0,1){100}}
\multiput(5,0)(21.43,0){5}{\line(0,1){3}}
\multiput(5,0)(4.286,0){21}{\line(0,1){1.4}}
\multiput(0,20)(0,20){4}{\line(1,0){2}}
\multiput(0,4)(0,4){25}{\line(1,0){1.4}}
\put(5,-8){\makebox(1,1)[b]{0}}
\put(26.,-8){\makebox(1,1)[b]{0.1}}
\put(47.,-8){\makebox(1,1)[b]{0.2}}
\put(68.,-8){\makebox(1,1)[b]{0.3}}
\put(90.,-8){\makebox(1,1)[b]{0.4}}
\put(-9,20){\makebox(1,.5)[l]{20}}
\put(-9,40){\makebox(1,.5)[l]{40}}
\put(-9,60){\makebox(1,.5)[l]{60}}
\put(-9,80){\makebox(1,.5)[l]{80}}

\put(75,90){\large $\Delta$I=90\%}
\put(5,105){\large F (\%)}
\put(75,-15){\large $t$ (s)}

\end{picture}
	\end{center}
	\caption{
Extraction efficiency measured with 0.65-mrad crystal ($\bullet$)
as a function of time during the spill,
for $\Delta$I=90\%.
Also shown are simulated absolute figures of efficiency (o)
in the assumption of uncompensated angular drift.
}
\end{figure}
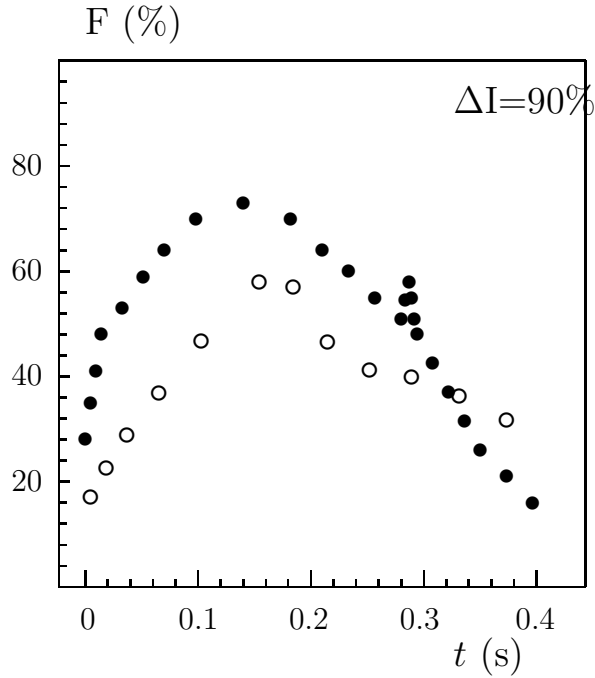

\section{Crystal Collimation Experiment}

Bent crystal, situated in the halo of a circulating beam,
can be the primary element in a scraping system,
thus serving as an 'active' collimator.
In this case, the only difference from extraction is that channeled particles
are bent onto a secondary collimator instead of the extraction beamline.
The bent particles are then intercepted (with
sufficiently big impact parameter) at the secondary element and
absorbed there.

We have performed the first demonstration experiment
on crystal-assisted collimation.
A bent crystal, with the same dimensions as the extraction crystals
described above and with bending angle of 1 mrad, was positioned
upstream of a secondary collimator
(stainless steel absorber 4 cm wide, 18 cm high, 250 cm long) "FEP"
and closer to the beam in the horizontal plane.
As the horizontal betatron tune is 9.73 in our accelerator,
it was most convenient to intercept the bent beam at FEP
not immediately on the first turn, but after 3 turns in the accelerator.
In this case the deflection angle of 1 mrad transforms into
more than 20 mm horizontal offset, and so the bent beam enters
the FEP collimator at some $\sim$15 mm from the FEP edge.
The optimal horizontal position of the crystal w.r.t. the FEP edge
was found to be $\sim$10 mm.

The radiation level was monitored at 3 places along the ring
in the vicinity of FEP.
Figure 3    shows how the radiation level depends on the angular
alignment of the crystal.
At the best crystal angle, preferable for channeling,
the radiation levels decrease by up to factor of $\sim$two in
the  places of monitoring.
This is explained by the fact that $\sim$50\% of the incident beam
is channeled by the crystal and deflected to the depth of FEP
where absorbed.
In the case when crystal was out and the beam was scraped
directly by FEP,
the radiation at the monitors was at about
the same level as in the case of disaligned crystal.

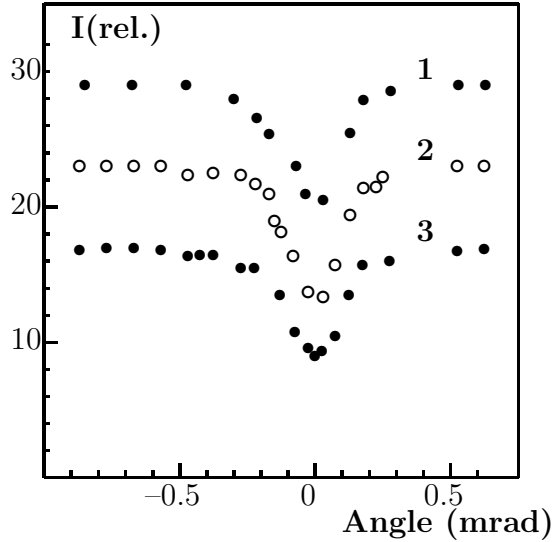
\begin{figure}[htb]
\begin{center}
\setlength{\unitlength}{1.8mm}
\begin{picture}(35,35)(0,-3)
\thicklines
\linethickness{.3mm}

\put(    27.5, 29.5)  {\bf 1}
\put(    32.6, 29)  {\circle*{.8}}
\put(    30.6, 29)  {\circle*{.8}}
\put(    25.6, 28.6)  {\circle*{.8}}
\put(    23.6, 27.9)  {\circle*{.8}}
\put(    22.6, 25.5)  {\circle*{.8}}
\put(    20.6, 20.5)  {\circle*{.8}}
\put(    19.3, 21)  {\circle*{.8}}
\put(    18.6, 23)  {\circle*{.8}}
\put(    16.6, 25.4)  {\circle*{.8}}
\put(    15.7, 26.6)  {\circle*{.8}}
\put(    14, 28)  {\circle*{.8}}
\put(    10.5, 29)  {\circle*{.8}}
\put(    6.5, 29)  {\circle*{.8}}
\put(    3, 29)  {\circle*{.8}}

\put(    27.5, 17.5)  {\bf 3}
\put(    32.5, 16.9)  {\circle*{.8}}
\put(    30.5, 16.75)  {\circle*{.8}}
\put(    25.5, 16)  {\circle*{.8}}
\put(    23.5, 15.7)  {\circle*{.8}}
\put(    22.5, 13.5)  {\circle*{.8}}
\put(    21.5, 10.5)  {\circle*{.8}}
\put(    20.5, 9.4)  {\circle*{.8}}
\put(    20, 9.)  {\circle*{.8}}
\put(    19.5, 9.6)  {\circle*{.8}}
\put(    18.5, 10.8)  {\circle*{.8}}
\put(    17.4, 13.5)  {\circle*{.8}}
\put(    15.5, 15.5)  {\circle*{.8}}
\put(    14.5, 15.5)  {\circle*{.8}}
\put(    12.5, 16.5)  {\circle*{.8}}
\put(    11.5, 16.5)  {\circle*{.8}}
\put(    10.6, 16.4)  {\circle*{.8}}
\put(    8.6, 16.8)  {\circle*{.8}}
\put(    6.6, 17)  {\circle*{.8}}
\put(    4.6, 17)  {\circle*{.8}}
\put(    2.6, 16.8)  {\circle*{.8}}

\put(    27.5, 23.5)  {\bf 2}
\put(    32.5, 23)  {\circle{.8}}
\put(    30.5, 23)  {\circle{.8}}
\put(    25, 22.2)  {\circle{.8}}
\put(    24.5, 21.5)  {\circle{.8}}
\put(    23.6, 21.4)  {\circle{.8}}
\put(    22.6, 19.4)  {\circle{.8}}
\put(    21.5, 15.7)  {\circle{.8}}
\put(    20.6, 13.4)  {\circle{.8}}
\put(    19.5, 13.7)  {\circle{.8}}
\put(    18.4, 16.4)  {\circle{.8}}
\put(    17.5, 18.2)  {\circle{.8}}
\put(    17, 19)  {\circle{.8}}
\put(    16.6, 21)  {\circle{.8}}
\put(    15.6, 21.7)  {\circle{.8}}
\put(    14.5, 22.4)  {\circle{.8}}
\put(    12.5, 22.5)  {\circle{.8}}
\put(    10.6, 22.4)  {\circle{.8}}
\put(    8.6, 23)  {\circle{.8}}
\put(    6.6, 23)  {\circle{.8}}
\put(    4.6, 23)  {\circle{.8}}
\put(    2.6, 23)  {\circle{.8}}

\put(0,0) {\line(1,0){35}}
\put(0,0) {\line(0,1){35}}
\put(0,35) {\line(1,0){35}}
\put(35,0){\line(0,1){35}}
\multiput(0,0)(2,0){17}{\line(0,1){.5}}
\multiput(0,0)(10,0){4}{\line(0,1){1}}
\multiput(0,0)(0,10){3}{\line(1,0){1}}
\multiput(0,0)(0,2){17}{\line(1,0){.4}}
\put(9,-2){\makebox(1,1)[b]{--0.5}}
\put(19,-2){\makebox(1,1)[b]{0}}
\put(29,-2){\makebox(1,1)[b]{0.5}}
\put(-2.5,30){\makebox(1,.5)[l]{30}}
\put(-2.5,20){\makebox(1,.5)[l]{20}}
\put(-2.5,10){\makebox(1,.5)[l]{10}}

\put(2,32.5){\bf I(rel.)}
\put(22,-4){\bf Angle (mrad)}

\end{picture}
\end{center}
\caption{
Measured irradiation in detectors 1, 2, 3 as function of crystal angle.
}
  \label{col1}
\end{figure}

We were able to check the crystal efficiency figure by alternative
means, measuring the profile and intensity of the particles
incident at the FEP entry face. The channeled beam had a narrow
profile and was well distanced from the FEP edge,
as shows Figure 4  where this profile is shown in comparison
with the profile of the accelerator beam deflected onto FEP by a kicker
magnet. From comparison of the two profiles, from crystal and from kicker,
we again derived the crystal efficiency, which was found to be
about 50\%, in agreement with the radiation monitoring figures
and with the earlier shown figures of extraction efficiency
with crystal in straight section 106.

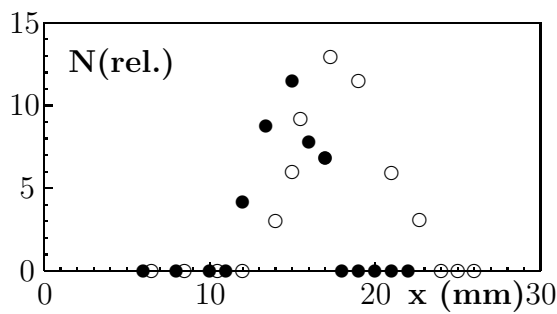
\begin{figure}[htb]
\begin{center}
\setlength{\unitlength}{.22mm}
\begin{picture}(300,150)(0,-3)

\put(    220.,0)  {\circle*{8}}
\put(    210.,0)  {\circle*{8}}
\put(    200.,0)  {\circle*{8}}
\put(    190.,0)  {\circle*{8}}
\put(    180.,0)  {\circle*{8}}
\put(    170.,68)  {\circle*{8}}
\put(    170.,68)  {\circle*{8}}
\put(    160.,78)  {\circle*{8}}
\put(    150.,115)  {\circle*{8}}
\put(    134.,88)  {\circle*{8}}
\put(    120.,42)  {\circle*{8}}
\put(    110.,0)  {\circle*{8}}
\put(    100.,0)  {\circle*{8}}
\put(    80.,0)  {\circle*{8}}
\put(    60.,0)  {\circle*{8}}

\put(    260.,0)  {\circle{8}}
\put(    250.,0)  {\circle{8}}
\put(    240.,0)  {\circle{8}}
\put(    227.,31)  {\circle{8}}
\put(    210.,59)  {\circle{8}}
\put(    190.,115)  {\circle{8}}
\put(    173.,129)  {\circle{8}}
\put(    155.,92)  {\circle{8}}
\put(    150.,60)  {\circle{8}}
\put(    140.,30)  {\circle{8}}
\put(    120.,0)  {\circle{8}}
\put(    105.,0)  {\circle{8}}
\put(    85.,0)  {\circle{8}}
\put(    65.,0)  {\circle{8}}

\put(0,0) {\line(1,0){300}}
\put(0,0) {\line(0,1){150}}
\put(0,150) {\line(1,0){300}}
\put(300,0){\line(0,1){150}}
\multiput(0,0)(100,0){3}{\line(0,1){5}}
\multiput(0,0)(10,0){30}{\line(0,1){2}}
\multiput(0,50)(0,50){3}{\line(1,0){5}}
\multiput(0,0)(0,10){15}{\line(1,0){2}}
\put(0,-20){\makebox(1,1)[b]{0}}
\put(100,-20){\makebox(1,1)[b]{10}}
\put(200,-20){\makebox(1,1)[b]{20}}
\put(300,-20){\makebox(1,1)[b]{30}}
\put(-21,0){\makebox(1,.5)[l]{ 0}}
\put(-21,50){\makebox(1,.5)[l]{ 5}}
\put(-21,100){\makebox(1,.5)[l]{10}}
\put(-21,150){\makebox(1,.5)[l]{15}}

\put(15,120){\bf N(rel.)}
\put(220,-20){\bf x (mm)}

\end{picture}
\end{center}
\caption{
Profiles measured at FEP entry face:
channeled beam ($\bullet$)
and beam (o) deflected by kicker magnet.
}
  \label{col2}
\end{figure}

\section{Conclusions}

The crystal-assisted method of slow extraction demonstrates
peak efficiency in the order of 60-70\%, now comparable to alternative
techniques of slow extraction,
and shows reliable, reproducible and predictable work.
A crystal can channel at least 5-6$\times$10$^{11}$ ppp
without cooling measures taken and without degradation seen.
The basis for the efficiency boost was the multiple character
of the particle encounters with a short crystal.

In our experiment
this technique was for the first time demonstrated for
scraping of the beam halo. Such application has been studied
by computer simulation for several machines,
notably RHIC \cite{rhic} and Tevatron \cite{tev}.
We have shown that radiation levels in accelerator can be
significantly decreased by means of channeling crystal incorporated
into beam cleaning system as a primary element.

We plan to continue tests with crystals 2.5-3.0 mm long,
where Monte Carlo predicts $\sim$80\% extraction efficiency.
The ultimate efficiency in our 70 GeV experiment
can be obtained with a crystal shortened to $\sim$1 mm.
We study different techniques to prepare bent crystal lattices with
required size, one of the most interesting approaches is described
in Ref.\cite{breese}.

\end{document}